\documentclass[preprint2, 12pt]{aastex62}
\usepackage[T1]{fontenc}

\usepackage{bm}        
\usepackage{amssymb, amsmath}   
\usepackage{cancel}
\usepackage{color}

\renewcommand{\thefootnote}{\arabic{footnote}}

\def\solphys{Sol.~Phys.}      
\def\apj{Astrophysical Journal}                 
\def\apjl{Astrophysical Journal}                 
\def\mnras{MNRAS}                 
\def\ssr{Space Science Reviews}                 
\def\aap{Astronomy and Astrophysics}                 

\begin{document}



\title{Determining the Dominant Acceleration Mechanism during Relativistic Magnetic Reconnection in Large-scale Systems}
\author{Fan Guo}
\affiliation{Los Alamos National Laboratory, NM 87545 USA}
\author{Xiaocan Li}
\affiliation{Los Alamos National Laboratory, NM 87545 USA}
\author{William Daughton}
\affiliation{Los Alamos National Laboratory, NM 87545 USA}
\author{Patrick Kilian}
\affiliation{Los Alamos National Laboratory, NM 87545 USA}
\author{Hui Li}
\affiliation{Los Alamos National Laboratory, NM 87545 USA}
\author{Yi-Hsin Liu}
\affiliation{Dartmouth College, Hanover, NH 03750 USA}
\author{Wangcheng Yan}
\affiliation{The University of Tennessee, Knoxville, TN 37996 USA}
\author{Dylan Ma}
\affiliation{Los Alamos National Laboratory, NM 87545 USA}
\date{\today}

\begin{abstract}
While a growing body of research indicates that relativistic magnetic reconnection is a prodigious source of particle acceleration in high-energy astrophysical systems, the dominant acceleration mechanism remains controversial. Using a combination of fully kinetic simulations and theoretical analysis, we demonstrate that Fermi-type acceleration within the large-scale motional electric fields dominates over direct acceleration from non-ideal electric fields within small-scale diffusion regions. This result has profound implications for modeling particle acceleration in large-scale astrophysical problems, since it opens up the possiblity of modeling the energetic spectra without resolving microscopic diffusion regions.
\end{abstract}



\section{Introduction} Magnetic reconnection is a plasma process that rapidly unleashes energy stored in magnetic shear into various forms of particle kinetic energy. It has been discussed in solar and space environments \citep{Kopp1976,Phan2000}, laboratory experiments \citep{Ji1998}, and recently in the context of high-energy astrophysics
\citep{Pino2005,Giannios2009,Arons2012,Hoshino2012,Zhang2011,McKinney2012,Zhang2015,Zhang2018}. There is strong observational evidence suggesting that reconnection is an efficient process for producing energetic particles in various heliophysics and astrophysical systems \citep{Birn2012,Krucker2010,Oieroset2002,Oka2018,Gary2018,Abdo2011,Tavani2011}. However, the acceleration physics remains an area of active research.

Recently, a growing body of research indicates that relativistic magnetic reconnection 
in the magnetically dominated regime
(magnetization parameter $\sigma=B^2/(4\pi n m c^2)\gg1$) is a prodigious source of high-energy particles in various astrophysical systems. However, the dominant acceleration mechanism remains controversial. Two identified candidates are direct acceleration at diffusion regions surrounding X-lines \citep{Zenitani2001,Pritchett2006,Fu2006,Uzdensky2011,Cerutti2013,Sironi2014,Wang2016} and Fermi-type acceleration within the much larger-scale reconnection layer \citep{Drake2006,Oka2010,Bessho2012,Dahlin2014,Guo2014,Guo2015,Li2017,Li2018a}. An additional controversy is the roles of the two mechanisms in producing the power-law particle energy distribution \citep{Sironi2014,Guo2014,Guo2015,Werner2016}. 
\citet{Sironi2014} have proposed that the power-law shape 
is established as the particles interact with the X-points (more specifically, diffusion regions with weak magnetic field $|E|>|B|$) through direct acceleration \citep{Zenitani2001}. They argue that this process is essential for the formation of power-law distributions and it determines the spectral index of the energy spectra. In contrast, \citet{Guo2014,Guo2015} proposed that the power-law distributions are produced by a Fermi-like process and continuous injection from the reconnection inflow. Based on this idea, they developed a theoretical model that is consistent with the hard spectra $f \propto \varepsilon^{-p}$ observed in simulations (approaching $p=1$,  where $p$ in the power-law index).

Determining the dominant acceleration mechanism and formation of power-law distribution in kinetic simulations is vital for understanding and building reconnection acceleration models for large-scale applications. Because of the enormous scale separation between the system size and the skin depth scale ($L/\lambda_e\sim10^8$ for solar flares, $\sim10^{13}$ for PWNe, and $\sim10^{17}$ for extragalactic jets)\citep{Ji2011}, it is impractical for conventional kinetic simulation methods to model the whole problem.
There have been attempts for modeling particle acceleration during magnetic reconnection in a macroscopic system by neglecting acceleration due to the non-ideal electric field \citep{Li2018b,Beresnyak2016,LeRoux2015,Drake2018}. To assess these models, it is important to determine if the non-ideal electric field plays any significant role in large-scale reconnection acceleration. The focus of this Letter is to distinguish the acceleration mechanisms and their roles in the development of the nonthermal power-law spectrum.  Through a combination of fully kinetic simulations and theoretical analysis, we demonstrate that the dominant acceleration mechanism is a Fermi-type process in the motional electric field induced from plasma motion in the reconnection layer. While the non-ideal electric field may act
as an additional particle injection for further Fermi acceleration, it is not necessary for the formation of power-law distributions and therefore can be neglected when modeling the energetic particle spectra in large-scale astrophysical reconnection events. 

\section{Numerical simulations} Simulations start from a force-free current layer with $\textbf{B}=B_0\text{tanh} (z/\lambda)\hat{x}+B_0\text{sech}(z/\lambda)\hat{y}$, corresponding to a magnetic field rotating by $180^\circ$ across the sheet. 
 The plasma consists of 
electron-positron pairs (mass ratio $m_p/m_e=1$). The initial distributions 
are Maxwellian with a uniform density $n_0$ 
and temperature ($T_{p}=T_{e}$). For the simulations presented here, the thermal energy per particle is $0.36m_ec^2$, but we have verified that our main conclusion is valid even when $T_p$ is as low as $0.01m_ec^2$.   
Particles in the sheet have a drift velocity $\textbf{u}_p=-\textbf{u}_e$, and that gives rise to a current density satisfying Ampere's law $\nabla\times\textbf{B}=4\pi\textbf{J}$. The simulations are 
performed using the VPIC \citep{Bowers2008} and NPIC codes \citep{Daughton2006,Daughton2007}, both of 
which solve the relativistic Vlasov-Maxwell equations but use different methods for solving the equations. We focus on the case with $\sigma_e = B^2/(4\pi n_e m_e c^2) = 100$ ($\sigma = 50$ including both species), corresponding to $\omega_{pe}/\Omega_{ce}=0.1$, where $\omega_{pe}$ is the plasma frequency and $\Omega_{ce}$ is the electron gyrofrequency.  Results for different $\sigma$ and domain size will be published elsewhere. The electric and magnetic fields are normalized by $B_0$. The domain size is $L_x\times L_z =600d_e\times 400d_e$, where $d_e=c/\omega_{pe} = c / (4 \pi n_e e^2/m_e)^{1/2}$ is the inertial length (without relativistic correction). The resolution of the simulations is $N_x\times N_z=3072\times2048$. All simulations used more than 100 particles per species per cell for each species, employed periodic boundary conditions in the $x$-directions, and in the $z$-direction used conducting boundaries for the fields and reflecting boundaries for the particles. The half-thickness is $\lambda=6d_e$. A small long-wavelength perturbation is included to initiate reconnection.

In VPIC simulations, we have developed a particle tracing module to output particle trajectories and find the electric field, magnetic field, and bulk fluid velocity at particle locations for studying particle energization \citep{Guo2016,Li2018a,Li2019}. In this study, we uniformly select $\sim1$ million particles in the beginning of the simulation and analyze their acceleration to high energy. In order to definitely demonstrate the acceleration physics, we developed the capability of including test particles that interact with magnetic fields in the normal manner, but only interact with the motional electric field $\textbf{E}_m=-\textbf{u}\times\textbf{B}/c$,  and do not experience any non-ideal electric fields. 
Note that because this technique requires us to calculate plasma flow velocity \textbf{u} from a finite number of particles, it introduces additional numerical noise to the test-particle component. For these simulations we use more self-consistent particles $1200$ per cell per species in the initial setup.
We also tag particles when they reached a region with weak magnetic field $|E|>|B|$, which is emphasized by \citet{Sironi2014}. 
Our earlier studies have shown that VPIC and NPIC give consistent results and we present results from the two codes without distinction.

\section{Distinguishing the Acceleration Mechanisms}  We attempt to distinguish two types of processes: The Fermi-type acceleration process in reconnection-driven bulk flows and direct acceleration in diffusion regions. While the Fermi-type acceleration is accomplished in the electric field induced by bulk plasma motion $\textbf{E}_m=-\textbf{u}\times\textbf{B}/c$, the non-ideal electric field that is associated with direct acceleration can be distinguished by the generalized Ohm's law \citep{Hesse2007,Liu2015,Bessho2005,Swisdak2008}

\begin{eqnarray}
\textbf{E}  &=& - \frac{\textbf{u} \times \textbf{B}}{c} + 
 \frac{1}{ne} \nabla \cdot (\textbf{P}_p-\textbf{P}_e) \\
 &+& \frac{m_e}{ne} \left[n_p (\frac{\partial \textbf{w}_p}{\partial t} + \textbf{u}_p \cdot \nabla \textbf{w}_p) 
 - n_e(\frac{\partial\textbf{w}_e}{\partial t} +  \textbf{u}_e \cdot \nabla \textbf{w}_e) \right],    \nonumber
\end{eqnarray}
where $n = n_p + n_e$ and we have assumed a pair plasma $m_p = m_e$ so the Hall term vanishes. $\textbf{P}_p $ and  $\textbf{P}_e$ are pressure tensors for the two particle species. $\textbf{w}_p$ and $\textbf{w}_e$ are moments of the spacelike components of the four velocity for each species, respectively \citep{Hesse2007,Liu2015}. Different from some earlier analysis \citep{Bessho2005,Swisdak2008}, the charge neutrality assumption  is dropped as local charge separation during relativistic reconnection can be quite large.

Based on Eq. (1), for each tracer particle one can distinguish the Fermi-like acceleration by calculating energy gain $\Delta\varepsilon_m=\int q\textbf{v}\cdot\textbf{E}_mdt$, where $\textbf{v}$ is the particle velocity, from the acceleration by the non-ideal electric field $\Delta \varepsilon_n=\int q\textbf{v}\cdot\textbf{E}_ndt=\int q\textbf{v}\cdot(\textbf{E}+\textbf{u}\times\textbf{B}/c)dt$ including the direct acceleration at X-line regions. In addition, we tag particles that entered diffusion regions with a strong electric field and weak magnetic field $|E|>|B|$ at least once, and calculate their energy distributions in the diffusion region and how they evolve elsewhere in the simulation domain. 

\section{Simulation Results} Fig. 1 contrasts y-components of motional electric field $E_{my}$ and non-ideal electric field $E_{ny}$ in the simulation at $\omega_{pe}t=400$. To better illustrate the fine-scale structure of the non-ideal electric field, both panels are magnified to the region $200<x/d_e<400$ and $-40<z/d_e<40$.  The motional electric field is primarily associated with the plasmoid motion and reconnection outflow. The non-ideal electric field is typically $\sim 0.1B_0$ \citep{Liu2017,Liu2015,Guo2015} and the motional electric field is typically $5-10$ times stronger compared to the non-ideal electric field. The non-ideal electric field is also present in the island region because of the non-zero divergence of pressure tensors in Eq. (1). In the rest of the Letter we discuss the relative roles of motional electric field and non-ideal electric field in accelerating particles.
\begin{figure}
\includegraphics[width=0.5\textwidth]{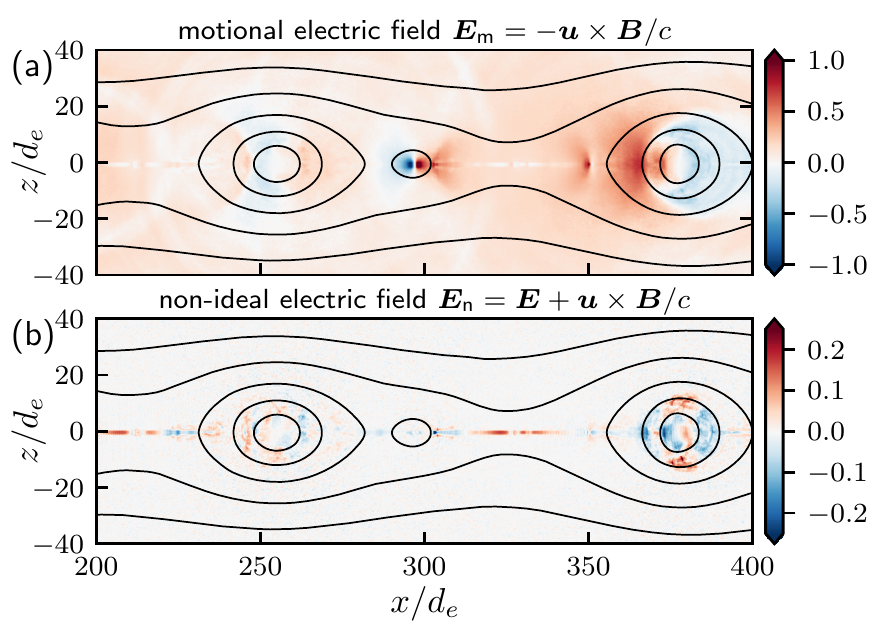}
\caption{The distribution of y-components of (a) motional electric field and (b) non-ideal electric field normalized by $B_0$ at $\omega_{pe}t=400$  \label{fig:epsart} overlaid by the contours representing magnetic field lines.}
\end{figure}

If the non-ideal electric field is essential for nonthermal acceleration in the reconnection layer, particles without significant direct acceleration would not be accelerated to high energy. We find that, however, while some high-energy particles experience an initial acceleration in the diffusion region with $|E|>|B|$, this process is not necessary because a significant number of high-energy particles did not pass through such regions (see below for more detailed discussions). 
Fig. 2(a) shows the trajectory of a particle in the energy gain versus $x$ plot. The blue line represents the energy gain, the orange line shows the contribution from
the motional electric field $\Delta \varepsilon_m$ and the green line indicates the contribution from the non-ideal electric field $\Delta\varepsilon_n$. This particle does not experience any significant non-ideal electric field acceleration and $\Delta\varepsilon_m$ dominates the energy increase (actually $\Delta\varepsilon_n<0$ most of the time). However, the particle still gains a dramatic amount of energy and eventually reaches $\gamma\sim 600$. 
Meanwhile, we use $\sim 1$ million tracer particles and track their energy evolution. We calculate the contributions from the motional electric field $\Delta \varepsilon_m$ and non-ideal electric field $\Delta \varepsilon_n$ for each particle during the acceleration process. Fig. 2(b) shows the averaged fractions of the energy gains from $\left<\Delta \varepsilon_m\right>$ (orange), $\left<\Delta \varepsilon_n\right>$ (green), and the region with $|E|>|B|$ (red) as a function of energy gain until the end of simulation. The contributions from motional and non-ideal electric fields are comparable at low energies, but Fermi-type acceleration becomes dominant when it accelerates particles to high energy, whereas the role of the non-ideal electric field is negligible. The effect of regions with $|E|>|B|$ is even smaller. This clearly demonstrates that the Fermi acceleration is the dominant mechanism for particle acceleration to high energy. 

\begin{figure}
\includegraphics[width=0.5\textwidth]{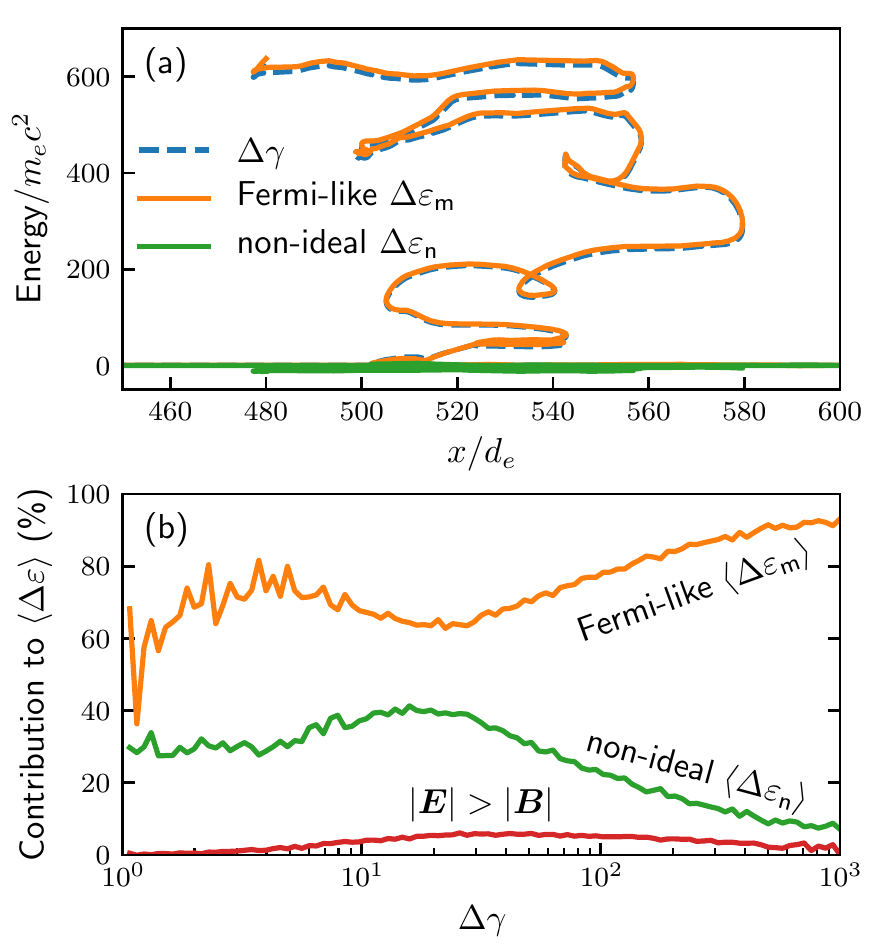}
\caption{Panel (a) shows a sample particle accelerated with Fermi-type acceleration dominating over non-ideal electric field acceleration. The Fermi acceleration does not rely on initial direct acceleration. Panel (b) shows statistics of energy gain for $\sim 1$ million particles traced over the history of the simulation. The orange line shows the fraction of averaged energy gain from motional electric field as a function of energy gain until the end of simulation. The green line shows the contribution of the non-ideal electric field and the red line shows the contribution of electric field in regions with $|E|>|B|$. The acceleration to high-energy is dominated by the Fermi-type acceleration process. \label{fig:epsart} }
\end{figure}

In Figure 3, we further examine the roles of the non-ideal electric field and motional electric field in forming the power-law distribution. Figure 3(a) shows the energy spectrum for particles in all of the diffusion regions with a strong non-ideal electric field and weak magnetic field $|E|>|B|$ and the energy spectrum integrated over the whole domain at $\omega_{pe}t=800$, respectively. Although quite variable, the representative energy spectrum in the diffusion region is nonthermal with a small spectral index $p\sim 0.4-0.5$ and an exponential cut off around $\gamma\sim 10-20$. Meanwhile, the spectral index for the energy spectrum over the whole domain is approximately $p=1.4$, consistent with previous works \citep{Sironi2014,Guo2014,Guo2015,Werner2016}. While previous study has claimed that the energy spectra in the diffusion region and the whole reconnection domain are the same \citep{Sironi2014}, the analysis here shows a clear difference. To understand this difference, we examine the energy continuity equation with injection  and escape of particles \citep{Zenitani2001,Drury2012,Guo2014,Guo2015}:
\begin{eqnarray}
\frac{\partial f}{\partial t}+\frac{\partial}{\partial \varepsilon} \left( \frac{\partial \varepsilon}{\partial t} f \right) = \frac{f_{inj}}{\tau_{inj}} - \frac{f}{\tau_{esc}},
\end{eqnarray}
where $\tau_{inj}$ and $\tau_{esc}$ are the injection and escape time scales for particles. 
We assume that some particles are pre-accelerated at X-points to a hard power-law with an exponential cutoff $f_{inj}=f_0(\varepsilon/\varepsilon_0)^{-\delta} \exp{(-\varepsilon/\varepsilon_c)}$, where $\delta=0.4$ is the spectral index, $\varepsilon_0=m_ec^2$ and the cut-off energy is $\varepsilon_c=10m_ec^2$ based on Fig. 3(a). Those particles are injected into island regions where particles are further accelerated by a Fermi-type process with acceleration rate $\alpha =\dot{\varepsilon}/\varepsilon$. The solution to (2) can be found, by integrating Eq.(2) along the characteristics from $t=0 $ to $t=\tau$ \citep{Drury1999,Guo2014}: 
\begin{eqnarray}
f(\tau_{}) = \frac{f_0 \varepsilon_c^{\theta}}{\alpha \tau_{inj} }[\Gamma_{\theta}(b)-\Gamma_{\theta}(be^{\alpha \tau)}]\varepsilon^{-(1+\beta)},
\end{eqnarray}
where $\beta=1/(\alpha \tau_{esc})$, $\theta=1+\delta-\beta$, $b=\varepsilon_0/\varepsilon_c$,
and $\Gamma_s(x)$ is the upper incomplete Gamma function.
In the limit of large $\alpha\tau$ (strong acceleration), the resulting energy spectrum is a power-law $f\propto\varepsilon^{-p}$ with $p=1+\beta$ in energy larger than the injected energy $\varepsilon_c$. We emphasize here that for generating a power-law distribution, the injected distribution does not have to be nonthermal \citep{Guo2014} and the actual value of $\delta$ does not alter the resulting spectral index. The value of $\alpha\tau$ can be estimated in PIC simulations \citep[e.g., as described in ][]{Guo2014} and for our simulations we obtained that 
$\alpha\tau=\int^{\tau_{inj}}_0\alpha dt\sim 4$, where we have assumed that the injection time $\tau_{inj}$ lasts until the saturation of reconnection. In Fig. 3(b) we plot Eq. (3) for $\alpha \tau=4,6,$ and $8$ 
and $\tau_{esc}\rightarrow\infty$, as our simulations do not include particle escape from the domain. 
This model predicts that the spectrum is steepened by the Fermi acceleration with $1<p<2$, consistent with PIC simulation results. For the strong acceleration case ($\alpha\tau=8$), the spectral index approaches $p = 1$, consistent with simulations for $\sigma \gg 1$ \citep{Sironi2014,Guo2014,Guo2015,Werner2016}.
These results suggest that Fermi acceleration does not require additional acceleration at X-points and the flat energy spectrum generated in X-lines is modified by Fermi acceleration within the outflow. We verify this by examining energy spectra for particles that never experience a diffusion region with $|E|>|B|$ and particles that did encounter at least one such region before $\omega_{pe}t=400$ (dashed lines) and $640$ (solid lines) in Fig. 3(c). At $\omega_{pe}t=400$, among the particles accelerated to $\gamma > 10$, only $25 \%$ of particles have encountered the regions with $|E|>|B|$. Particles that never encounter such a diffusion region still develop a clear power-law distribution. Later on, more particles went through the diffusion region but the spectral indices for these two classes of particles are still quite similar, confirming the basic conclusions from the analytical model. For sufficiently high energy ($\gamma \gtrsim 10$), the spectral indices for these two classes of particles are quite similar, confirming the basic conclusions from the analytic model.

\begin{figure}
\includegraphics[width=0.5\textwidth]{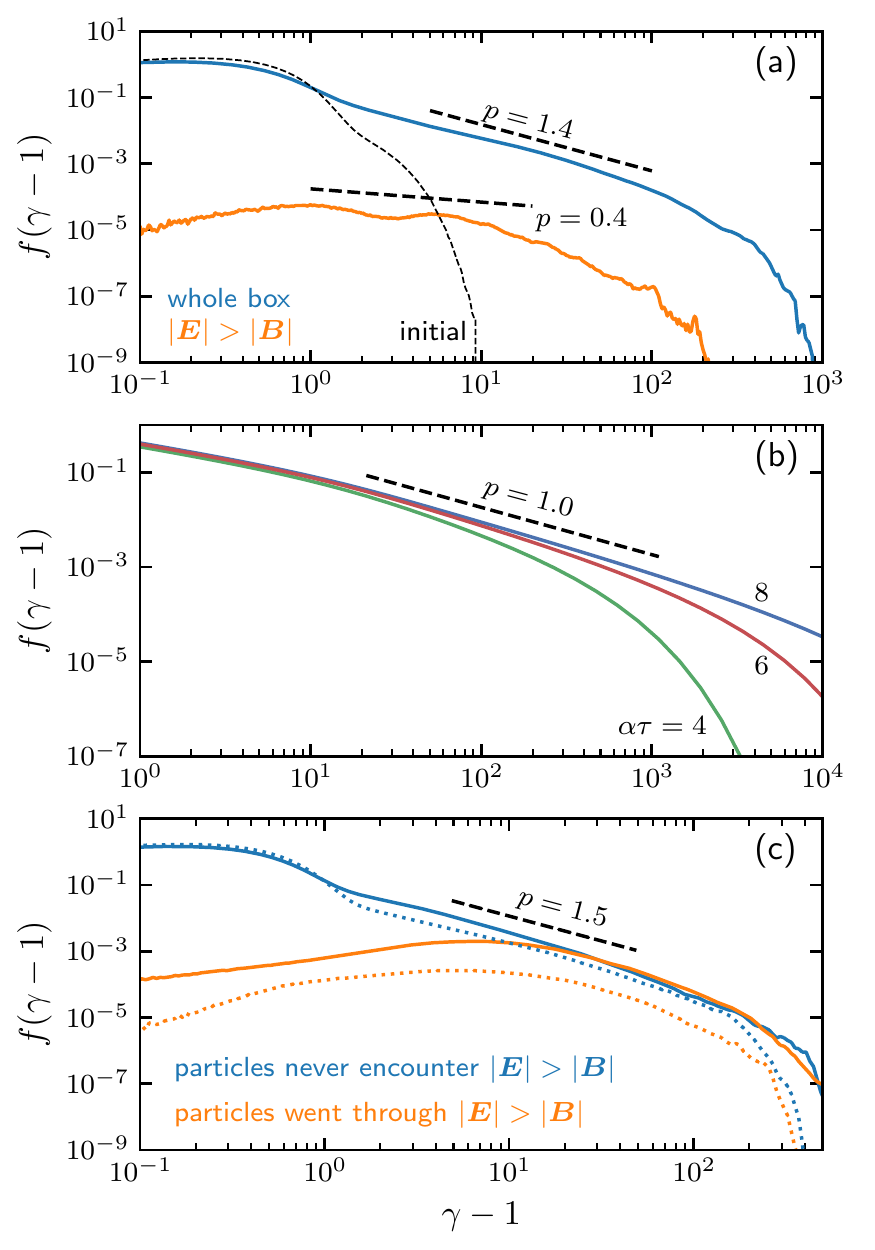}
\caption{(a) Energy spectra for particles over the whole domain (blue) and only the regions with $|E| > |B|$ (orange). The spectral indices for the two regions are significantly different from each other. (b) Illustration showing Eq. (3) for $\alpha \tau = 4, 6, $ and $8$ with $\delta = 0.4$, $\varepsilon_c = 10 m_ec^2$ and $\tau_{esc} \rightarrow \infty$. A flat injected energy spectrum is steepened to $1<p<2$ by Fermi acceleration. (c) Energy spectra for particles that never experienced the region (blue) with $|E|>|B|$ and particles that encountered at least one such region (orange) before $\omega_{pe}t=400$ (dashed lines) and 640 (solid lines). The two energy spectra give similar indices at high energy. \label{fig:epsart} }
\end{figure}

This analysis demonstrates that efficient Fermi acceleration does not require direct acceleration at X-lines. Furthermore, the energetic particles generated within the diffusion region are modified by the same Fermi-like process in the outflows. Ultimately, at high energy, the spectral indices are nearly the same, regardless of whether the particles ever encounter a diffusion region. This indicates that the non-thermal spectra resulting from relativistic magnetic reconnection can be computed by ignoring the influence of the non-ideal electric field.
 To directly demonstrate this, we performed an additional simulation with a test-particle electron component that does not feedback to the system. The test-particle electrons have the same initial distribution and one tenth the number of particles of electrons that is self-consistently evolved in the simulation, but only experience the motional electric field $\textbf{E}_m$ during the simulation. While these test-particles are not self-consistently evolved in the simulation, they do retain guiding-center drift motions such as electric field drift, gradient and curvature magnetic drift, and therefore can experience Fermi and betatron acceleration processes. This approach completely removes the acceleration associated with non-ideal electric field, but keeps Fermi-type acceleration in the reconnection layer. Fig. 4(a) shows the density of test-particle population and they are mainly concentrated in magnetic islands as expected. Fig. 4(b) shows the energy spectrum for the test-particle component. Test-particle electrons develop a power-law like energy spectrum with similar to self-consistent electrons in Fig. 3(a).
 The main extent of the power-law distribution is still preserved with a cut off energy $\gamma \sim 100$, indicating it includes the main physics necessary for developing power-law distributions.

\begin{figure}
\includegraphics[width=0.5\textwidth]{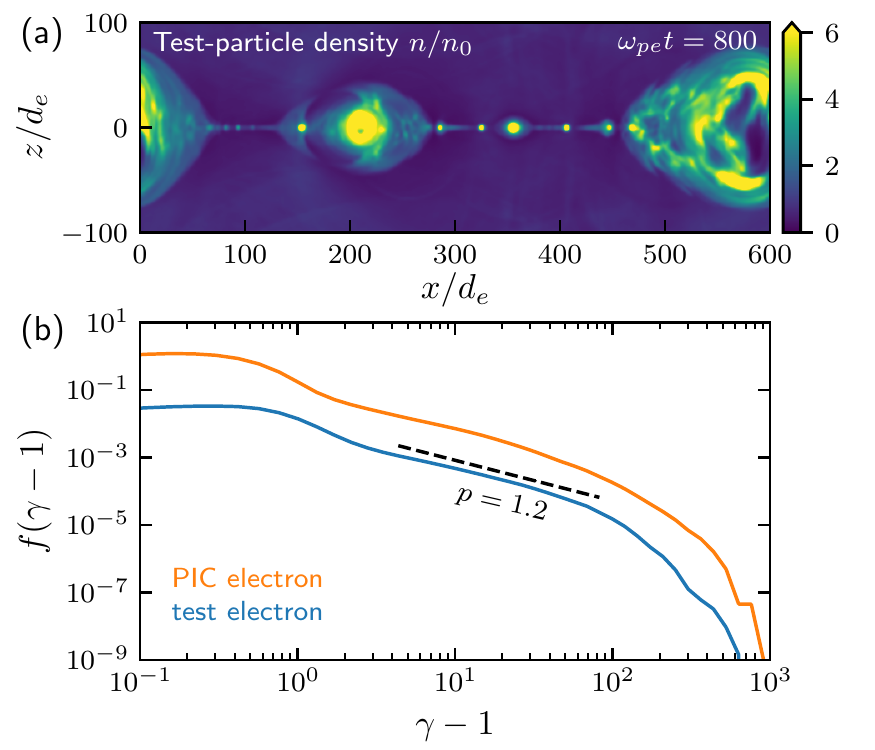}
\caption{The evolution of a test-particle population that does not experience non-ideal electric field in the PIC simulation at $\omega_{pe}t = 800$. Panel (a) shows the test-particle density in a snapshot. Panel (b) shows the energy spectrum for the test-particle population. A power-law distribution with a spectral index similar to that of self-consistent electrons is developed, even excluding the non-ideal electric field. \label{fig:epsart} }
\end{figure}

\textit{Discussion and Conclusion} --
The dominant acceleration mechanism in relativistic reconnection has been a controversial issue \citep{Sironi2014,Guo2014,Guo2015}.
Recently, \citet{Petropoulou2018} studied the long-term evolution of energy spectrum in large two-dimensional kinetic simulations of relativistic reconnection and found the break energy sustainably increases and spectrum continuously softens. The determining acceleration process they found is consistent with the current study and \citet{Guo2014,Guo2015}. In general, the evolution of spectral index can be studied using the analysis we proposed as well. However, we remark that effects like particle loss from more realistic boundary conditions as well as 3D effects would be important to consider. 

A major challenge for describing particle acceleration during magnetic reconnection in large-scale astrophysical system is the enormous scale separation between the system scales and plasma kientic scales.  Results of this Letter clearly demonstrate that the formation of power-law distributions does not rely on the non-ideal electric field. While the non-ideal electric field at X-points does accelerate a small population of particles, Fermi-type acceleration dominates over the direct acceleration and determines the spectral index. Therefore the X-line acceleration may be parameterized as an additional injection process for further Fermi-type acceleration.  For large-scale applications of reconnection, the non-ideal electric field is concentrated at boundary layers, and it may not be a significant source of energetic particles in a macroscopic reconnection event. This conclusion has profound implications for modeling particle acceleration in large-scale astrophysical systems, as it opens up the possibility of modeling the energetic particle spectra without resolving microscopic diffusion regions. 

\section{Acknowledgements}
We gratefully acknowledge discussions 
with Joel Dahlin, Jim Drake, Sasha Philippov, Lorenzo Sironi,  Anatoly Spitkovsky, and Marc Swisdak. We are grateful for support from DOE through the LDRD program at LANL and DoE/OFES support to LANL. X.L.'s contribution is in part supported by NASA under grant NNH16AC60I. The research by P. K. was also supported by the LANL through its Center for Space and Earth Science (CSES). CSES is funded by LANL's Laboratory Directed Research and Development (LDRD) program under project number 20180475DR. Simulations were performed at National Energy Research Scientific Computing Center (NERSC) and with LANL institutional computing.




\end{document}